\documentclass{osa-article}

\journal{osajournal}

\articletype{Research Article}

\usepackage{siunitx}
\begin{document}
\title{Nonlinear wavefront reconstruction with Convolutional Neural Networks for Fourier-based wavefront sensors}

\author{R. Landman\authormark{1,*}, S.Y. Haffert\authormark{1}}

\address{\authormark{1}{Leiden Observatory, Leiden University, PO Box 9513, Niels Bohrweg 2, 2300 RA Leiden, The Netherlands}}

\email{\authormark{*}rlandman@strw.leidenuniv.nl} 




\begin{abstract}
Fourier-based wavefront sensors, such as the Pyramid Wavefront Sensor (PWFS), are the current preference for high contrast imaging due to their high sensitivity. However, these wavefront sensors have intrinsic nonlinearities that constrain the range where conventional linear reconstruction methods can be used to accurately estimate the incoming wavefront aberrations. We propose to use Convolutional Neural Networks (CNNs) for the nonlinear reconstruction of the wavefront sensor measurements. It is demonstrated that a CNN can be used to accurately reconstruct the nonlinearities in both simulations and a lab implementation. We show that solely using a CNN for the reconstruction leads to suboptimal closed loop performance under simulated atmospheric turbulence. However, it is demonstrated that using a CNN to estimate the nonlinear error term on top of a linear model results in an improved effective dynamic range of a simulated adaptive optics system. The larger effective dynamic range results in a higher Strehl ratio under conditions where the nonlinear error is relevant. This will allow the current and future generation of large astronomical telescopes to work in a wider range of atmospheric conditions and therefore reduce costly downtime of such facilities.
\end{abstract}

\section{Introduction}\label{Introduction}
The past few decades have seen an increase in the use of adaptive optics (AO) to correct for wavefront errors in imaging systems. These wavefront errors are usually created when the light travels through turbulent media such as Earth's atmosphere in astronomy\cite{roddier1981}, or intracellular fluids in microscopy\cite{schwertner2004measurement}. Accurate measurements of the aberrations are necessary to reach the best image quality after compensation. In most systems a dedicated optical system, a wavefront sensor, is used to measure the aberrations because the phase of visible and near-infrared light can not be directly measured. A wavefront sensor transforms the incoming wavefront aberrations into measurable intensity modulations.

The reconstruction of the wavefront from these sensor measurements is not straightforward and almost all commonly used methods depend on a linearization of the response of the wavefront sensor around a non-aberrated wavefront \cite{Ellerbroek2009,Hutterer2018b,Heritier2018}. However, this assumption of linearity is not always valid as most wavefront sensors have intrinsic nonlinearities when the input wavefront has large deviations\cite{burvall2006linearity, guyon2009high, Haffert2016}. To reconstruct the wavefront accurately it is necessary to use nonlinear reconstruction (NLR) methods.

A particular focus of NLR in astronomy has been on the reconstruction of the unmodulated Pyramid Wavefront sensor (PWFS) \cite{ragazzoni1996pupil, ragazzoni2002pyramid} measurements. The PWFS is a promising wavefront sensor for astronomy due to its high sensitivity, which allows astronomers to use AO on fainter stars \cite{ragazzoni1999sensitivity,Verinaude2004}. The linear range of the unmodulated PWFS is relatively small and depends on the wavefront aberration mode. For tip and tilt errors the linear range is 1 rad while for higher-order modes it is even smaller\cite{burvall2006linearity, Fauvarque2016}. Atmospheric turbulence usually generates wavefront errors that are outside the linearity range of the unmodulated PWFS, which is why most, if not all, of the current AO systems add modulation to increase the linear range \cite{esposito2012lbt, males2016magaox, bond2018nirpwfs}. While the modulation does increase the linear range and allows the PWFS to be used in realistic conditions, the sensitivity also decreases \cite{Verinaude2004,guyon2005aowfs,Fauvarque2016}. 

As the PWFS response depends on the shape of the input Point Spread Function (PSF), there is also strong nonlinear cross-talk between different modes which quickly leads to an underestimation of the actual wavefront\cite{esposito2001pyramid,deo2018modal}. Recent work has shown that the use of model based nonlinear reconstruction methods can improve the reconstruction accuracy and increase the effective dynamic range of the PWFS \cite{Frazin2018,Hutterer2018}.


Other approaches to increase the linear range of the sensor focus on changing the hardware. As briefly discussed above, dynamic modulation of the input beam on top of the pyramid face can be used to increase the linear range, but this decreases the sensitivity\cite{ragazzoni1996pupil}. Different modulation schemes and even other types of focal plane masks instead of the pyramidal prism can be described in the general formalism of Fourier-based wavefront sensing \cite{Fauvarque2016}, where a two lens system is used with a focal plane mask in its intermediate focus to measure wavefront errors. This formalism shows that there is a linear trade-off between the sensitivity of the wavefront sensor and its linear range. The only way to get around this linear trade-off between sensitivity and dynamic range is to introduce nonlinearities in the wavefront sensor. This approach led to the development of the generalised Optical Differentiation Wavefront Sensor (g-ODWFS)\cite{Haffert2016}.

\begin{figure}[htbp]
    \centering
    \includegraphics[width=\linewidth]{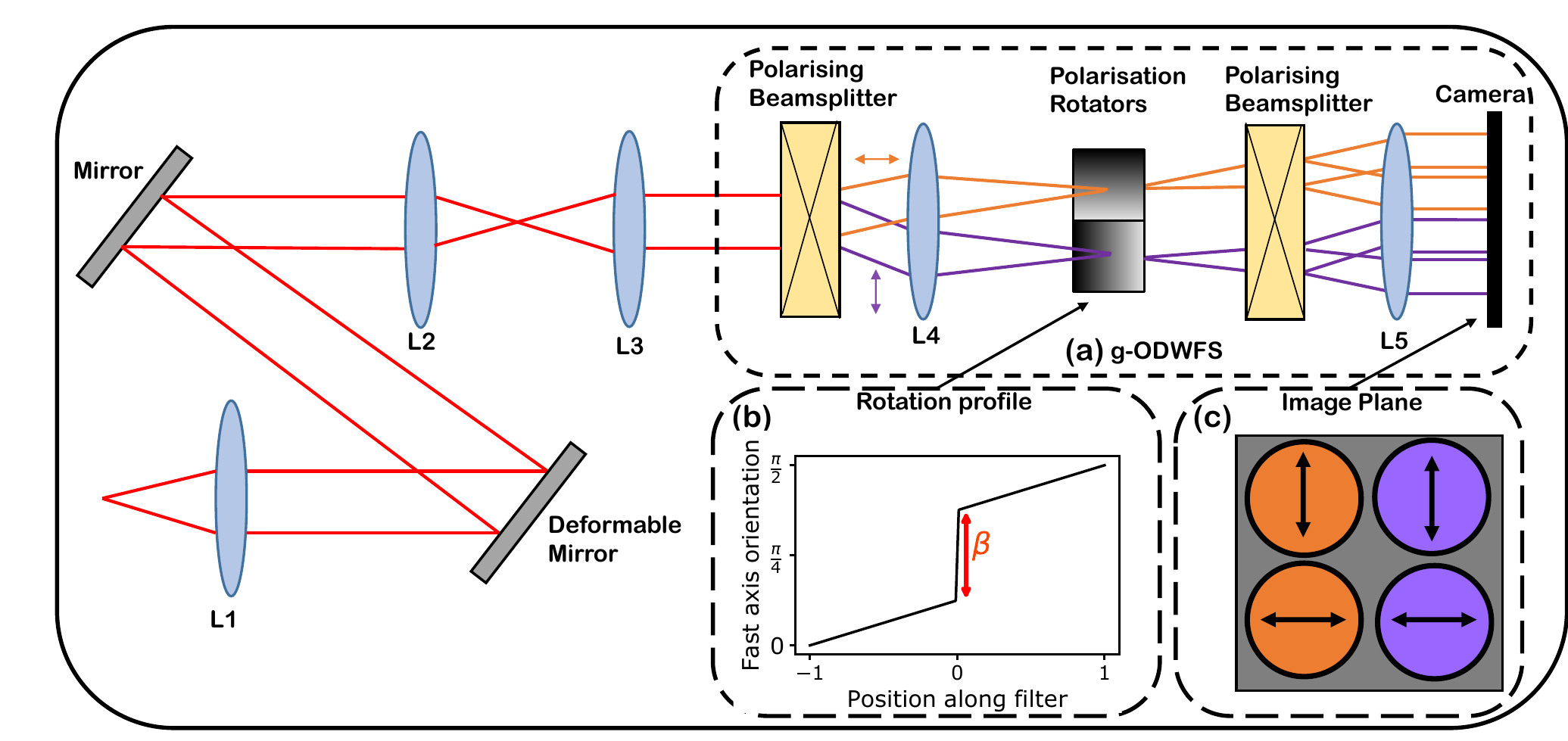}
    \caption{ \textbf{a)} Schematic illustration of the experimental setup of the generalized Optical Differentiation Wavefront Sensor. The incoming beam is collimated and reflected by a fold mirror onto a deformable mirror. A relay system is used to magnify the pupil, which is then split by a Wollaston prism into two beams. Each beam passes through a different polarisation filter in the focal plane to measure the wavefront gradient in the x and y direction. After the polarization filters a second Wollaston prism splits the beams again. The four output beams are collimated by the last lens. \textbf{(b)} Orientation of the fast axis for the half-wave plate rotators as function of position with a step size $\beta$. \textbf{c)} The aligment of the output pupils that encode the wavefront gradients.}
    \label{fig:godwfs}
\end{figure}

The g-ODWFS is similar to Fourier-based sensors but uses two filters in the focal plane instead of a single mask. Each filter encodes either the x-gradient or the y-gradient of the wavefront aberrations. A schematic of this setup can be seen in Figure \ref{fig:godwfs}. A Wollaston prism splits the collimated input beam into two polarized beams, which are focused onto the different focal plane filters. For the g-ODWFS the focal plane filters are  half-wave plates with a spatially varying fast-axis orientation, therefore the angle of polarization of the input beams will rotate by an amount that depends on its spatial position in the focal plane. A second Wollaston prim that is rotated by 45 degrees with respect to the first Wollaston prism is used as a polarization analyzer that creates four output beams, two for each gradient. The intensity difference between the two output beams per gradient depends on the angle of polarization, and therefore on the orientation of the fast axis of the focal plane mask where the beam hit it. We can measure for each position in the pupil what the angle of polarization is and derive where the ray hit the focal plane mask. From this we can measure the local tilt of the wavefront and reconstruct the wavefront aberrations. The fast-axis orientation profile is a step-wise profile parameterized by $\beta$ which determines how close the profile is to a step function. This profile can be seen in Fig. \ref{fig:godwfs} b.

Mathematically the g-ODWFS can be described as a Fourier filter,
\begin{equation}
    E_1 =\mathcal{F}\left[ T \mathcal{F}\left[E_0\right] \right].
\end{equation}
Here $E_0$ is the input electric field, $\mathcal{F}$ is the Fourier transform operator, $T$ is the focal plane transmission mask and $E_1$ is the output electric field. The g-ODWFS uses two tranmission masks, one that measures wavefront variations in the x direction and on that measures variations in the y direction. For a single direction, we choose to describe the x-direction, the transmission mask creates two effective amplitude filters,
\begin{equation}
    T_+ = \sin{\left[ \frac{\pi}{4}\left(\left(1+x\right)\left(1-\beta\right) + \beta H(x) \right) \right]},
\end{equation}
and,
\begin{equation}
    T_- = \cos{\left[ \frac{\pi}{4}\left(\left(1+x\right)\left(1-\beta\right) + \beta H(x) \right) \right]}.
\end{equation}
We create the two perfectly mirrored focal plane transmission masks with the sine and cosine, which come from the projection onto the two polarization states. The normalized difference between the corresponding pupils can then be used to measure the wavefront. In the definition of the filter masks, $x$ is used a normalized focal plane coordinate which runs from -1 to 1. Outside this range the filter has a transmission of zero. The power of the nonlinear step is defined by the interpolation parameter $\beta$ and the step itself is created by the heavyside function $H(x)$. The normalized difference in intensity between the two pupils, in the small phase approximation, is then proportional to,
\begin{equation}
    \frac{I_+ - I_-}{I_+ + I_-} = \left(1-\beta\right) k \frac{\partial W}{\partial x} + \beta P(\phi).
\end{equation}
Here $I_\pm$ is the pupil intensity of the corresponding transmission mask $T_\pm$, $k$ is the linear optical gain of the WFS \cite{Haffert2016} and $P$ is WFS response from the step function due to a wavefront abberation $\phi$. The measurement from a step filter is complicated and is a mixture of the wavefront gradient and the wavefront itself \cite{Verinaude2004, Hutterer2018}. The response from a step filter is very similar to the PWFS response. Some works even model the response of a PWFS by two step functions, because the step response is so similar in behaviour to the unmodulated PWFS \cite{Hutterer2018}.

The g-ODWFS has been demonstrated in the lab and recently at the telescope as a wavefront sensor that can be used for astronomical adaptive optics \cite{haffert2018sky}. Through the introduction of an intentional nonlinearity we can increase the dynamic range without decreasing the sensitivity of the wavefront sensor. However, for large wavefront errors the response is nonlinear, which shows that the only way to actually increase the effective dynamic range of these wavefront sensors is through nonlinear reconstruction of the incoming wavefront.

Instead of using a model based approach to nonlinear wavefront reconstruction, such as done by \cite{Hutterer2018,Frazin2018}, it is also possible to use a data-driven approach. The advantages of a data-driven approach is that it is not necessary to know the exact details of the optical system and its environment. Model based methods are very sensitive to any model errors, which can even dominate the reconstruction error \cite{Hutterer2018}. Deep learning is a promising approach to data-driven learning of complicated nonlinear relations between input-ouput sets. The use of neural networks has been explored in the past, but focused mainly on the comparison between the precision of linear reconstructors and neural networks \cite{Guo2006,AlvarezDiez2008}.

In this work we propose to use a Convolutional Neural Network (CNN) as reconstructor to extend the effective dynamic range of Fourier-based wavefront sensors well into the nonlinear regime where conventional linear reconstructors degrade substantially in quality. Section \ref{section:Reconstruction} describes the different reconstruction methods that are used in this paper. Section \ref{section:Simulations} provides a comparison between the reconstruction methods in simulations to show the potential gain in an idealised system. Section \ref{section:Lab} discusses the results from applying the CNN in a lab setup of the Leiden EXoplanet Instrument (LEXI)\cite{haffert2018sky} that has an implementation of the g-ODWFS as its main wavefront sensor.

\section{Reconstruction methods}\label{section:Reconstruction}
 The phase profile of the deformable mirror $\phi_{DM}(x,y)$ can be described by a set of modal functions $f_i(x,y)$:
 \begin{equation}
 \phi_{DM}(x,y) = \sum_i c_i f_i(x,y)
 \end{equation}
 Modal reconstruction involves the estimation of the modal coefficients $c_i$ from the slope measurements of the wavefront sensor. Throughout this paper, the actuator modes will be used as the modal basis. The actuator modes consist of the phase profiles induced when the individual actuators on the DM are poked. For the g-ODWFS, the wavefront slope along one direction is related to the normalized differences between the pupil intensities resulting from that polarisation rotator, as explained in the previous section. The normalized differences will be used as measurement vector throughout this work.
 
 \subsection{Matrix Vector Multiplication (MVM)}\label{section:MVM}
 The most common technique of wavefront reconstruction is to assume a linear relation between the measurement vector $\Vec{s}$ and the modal coefficients $\Vec{c}$. The forward model is then given by:
 \begin{equation}
     \Vec{s} = A\Vec{c} + \Vec{\eta}
 \end{equation}
 Where $A$ is the interaction matrix and $\Vec{\eta}$ some measurement noise from the WFS. The interaction matrix $A$ can be calibrated by measuring the actuator response within the linearity range of the wavefront sensor. The inverse operation of reconstructing the wavefront from the slope measurements involves the estimation of the least-squares estimate of $\hat{\Vec{c}}$. This estimation requires inversion of the interaction matrix. To avoid degradation of the reconstruction by small singular values, regularization can be added. This can for example be done by truncating the singular value decomposition, where small singular values are set to zero, or Tikhonov regularization. We will use Tikhonov regularization in the inversion, which gives the following cost function:
 \begin{equation}
     J = \|(A\hat{\Vec{c}}-\Vec{s})\|^2+\|\Gamma \hat{\Vec{c}}\|^2.
 \end{equation}
Where $\|\|$ denotes the $L^2$ norm and $\Gamma$ is the Tikhonov matrix. We choose $\Gamma = \alpha I$, with $I$ the identity matrix and $\alpha$ the regularization strength. The modal coefficients that minimize this cost function are given by:
 \begin{equation}
     \hat{\Vec{c}} = (A^T A + \alpha I)^{-1} A^T\Vec{s}.
 \end{equation}
 The reconstruction matrix $(A^TA + \alpha I)^{-1}A^T$ can be precomputed. The optimal value for the regularization strength $\alpha$ depends on the noise statistics and will be chosen by iteratively trying different values and choosing the value that gives the lowest mean squared error on a training set for the used noise level.

 \subsection{Convolutional Neural Networks (CNN)}
 Convolutional Neural Networks have proven to be excellent at various image processing and classification tasks \cite{Krizhevsky2012,He2015}. They use the assumptions that features are local and translational invariant to drastically reduce the number of weights that have to be optimized as compared to traditional Neural Networks. The main building blocks of CNNs are convolutional layers, in which kernels of weights are convolved with the image to produce feature maps. A CNN often consists of many stacked convolutional layers and a few Dense layers at the end, in which all inputs and outputs are connected. After each layer, a nonlinear activation function is applied, allowing it to model nonlinear relations. The model is learned by optimizing the weights of the CNN for a loss function on known input-output pairs. This is done by using the backpropagation algorithm in combination with a gradient based optimizer.
 
 For the reconstruction of the g-ODWFS measurements, the input of the CNN consists of the observed normalised differences in both directions, resulting in a 3D input array with a depth of 2. This input array is propagated through a number of convolutional layers. Finally, a single Dense layer is used to produce the output of the model, the estimated modal coefficients that are present in the wavefront. Using the actuator modes as modal basis effectively uses the assumptions from the convolutional layer, since all actuators have the same local response. This is in contrast to for example the Zernike basis, which depends on various global features. As the number of mirror modes and pupil dimensions are different in the simulations and lab work, the architectures are slightly different and the details will be discussed in their respective sections.

 The calibration is done by applying known combinations of the mirror modes on the DM and measuring the response. Using the Adam optimizer \cite{Kingma2014} we minimize a loss function for the observed measurements and applied modal coefficients. Commonly, the mean squared error is used as the loss function for regression problems. This is also the initial cost function that was tried to optimize the CNN, but it did not lead to satisfying results. For AO systems we need to provide gain over the full range of wavefronts that are observed in closed-loop, ranging from Strehls of $<0.1\%$ to Strehls of $>80\%$.  The mean squared error takes the square of the absolute error, which effectively causes the optimizer to focus more on wavefronts with relatively larger input RMS. This would not be a problem for linear regression, because a better solution for the large coefficients would also imply a better solution for the small coefficients due to linearity of the problem. For nonlinear regression with neural networks this does not hold, because neural networks can introduce non-continuous piecewise functions, which effectively means that the linear range and nonlinear range could be completely decoupled. This makes it easier for the optimizer to reduce the total cost by improving the fit in the large aberration regime while ignoring the small aberration regime. As this is not desirable for AO systems, we have opted to weigh the mean squared error by the square of the RMS of the true modal coefficients, leading to the following relative loss function:
 \begin{equation}
     J = \frac{\left<(c-\hat{c})^2\right>}{\left< c^2 \right>+\epsilon}
     \label{eq:loss}
 \end{equation}
 Here, $\left<\right>$ denotes the average along the modes and $\hat{c}$ denotes the predicted modal coefficients. The $\epsilon$ term avoids a diverging loss for small input RMS. This is an additional hyperparameter that can be tweaked. With this relative loss function we enforce the algorithm to put the same amount of weight on wavefronts with small RMS and those with large RMS. Our models are implemented and trained using the Keras package \cite{chollet2015keras} for Python.
 
 \subsection{CNN+MVM}
\begin{figure*}[htbp]
     \centering
     \includegraphics[width=\linewidth]{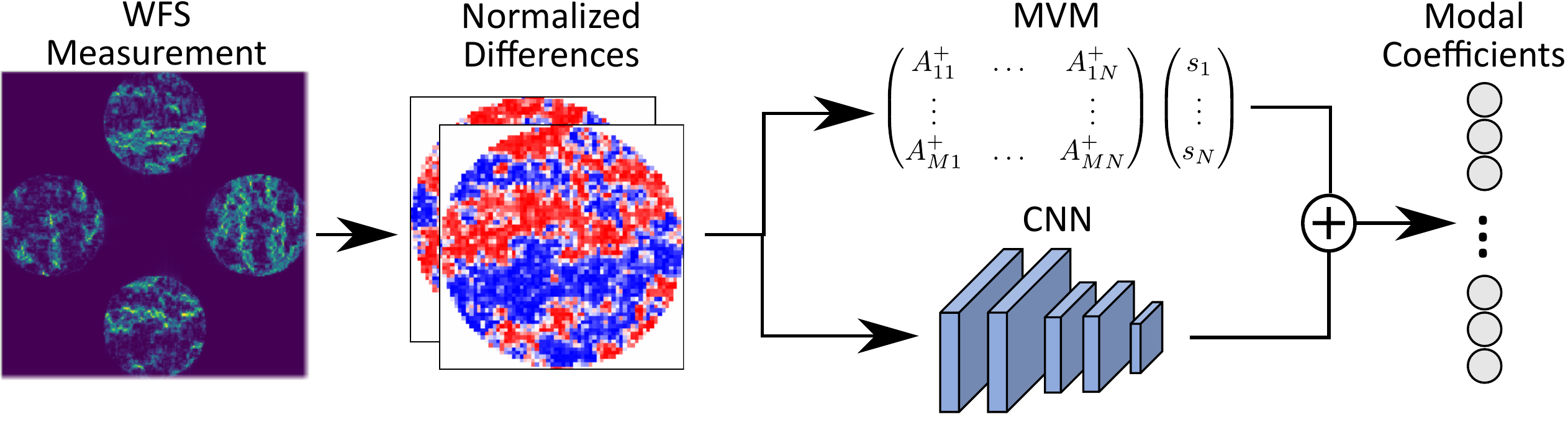}
     \caption{Overview of the pipeline for the nonlinear reconstruction of the wavefront for the CNN+MVM approach. The normalized differences are calculated from the WFS measurements. These normalized differences are then multiplied with the reconstruction matrix and propagated through the CNN. The results are summed and contain the estimated modal coefficients present in the incoming wavefront.}
     \label{fig:dataflow}
 \end{figure*}

 As will be shown in section \ref{section:closed_loop}, solely using a CNN results in suboptimal closed-loop performance under simulated atmospheric turbulence. We have solved this problem by decomposing the reconstruction in a linear term $\vec{c}_{l}$ and a nonlinear error term $\vec{c}_{nl}$:
 \begin{equation}
     \vec{c} = \vec{c}_l+\vec{c}_{nl}.
 \end{equation}
The MVM method discussed in the section \ref{section:MVM} can be used to estimate the linear term, while the CNN is calibrated to only reconstruct the nonlinear error term:
\begin{equation}
    \hat{\vec{c}} = A^{+}\vec{s} + \textrm{CNN}(\vec{s}).
\end{equation}
Where $A^+$ refers to the regularized inverted interaction matrix described in section \ref{section:MVM}. Figure \ref{fig:dataflow} shows the pipeline for reconstructing the wavefront sensor measurements for this approach. The implementation of the CNN is the same as in the previous section.
 
\section{Simulations}\label{section:Simulations}

\subsection{Simulation setup}
All simulations are done in Python with the hcipy \cite{por2018hcipy} package, which contains an implementation of the g-ODWFS. The deformable mirror has 32 actuators across the pupil diameter with a total number of 848 illuminated actuators over the full pupil. Each of the actuators has a Gaussian influence function with a standard deviation equal to the actuator pitch. Unless noted otherwise, we assume that the WFS measurements contain $10^6$ photons per frame. We will test the performance of the reconstruction method for three different step sizes $\beta$: 0, 1/3 and 1. The g-ODWFS with a step size of 1 is the equivalent of the unmodulated PWFS and is a good estimation of the performance that could be obtained with this WFS. The physical parameters used in the simulations are shown in Table \ref{tab:simulation_parameters}. 

\begin{table}[htbp]
\renewcommand{\arraystretch}{1.1}
\centering
\caption{\bf Simulation Parameters}
\begin{tabular}{lc}
\hline
Wavelength $\lambda$ & \SI{1}{\micro\metre}\\
Aperture diameter $D$ & 8 m \\
Filter size & 50 $\lambda/D$\\
Resolution input wavefront & 256x256 \\
Resolution output image & 64x64 per pupil\\
\end{tabular}
  \label{tab:simulation_parameters}
\end{table}

\subsection{Calibration}
The input array for the CNN, containing the normalized differences in both directions, has a shape of 64x64x2. We apply five layers of 3x3 convolutions with 32 kernels to this input array. The third convolutional layer uses a stride of two, reducing the spatial dimensions of the image. After these layers we have a single 1x1 convolution with 4 kernels to reduce the depth, which results in a 32x32x4 array. After each convolutional layer we apply a batch normalization layer \cite{Ioffe2015} and the ReLU activation. Finally, we flatten the array and have one fully connected layer with 848 neurons, equal to the number of mirror modes. This final layer introduces a lot of parameters, most of which are irrelevant because the actuator coefficients should mostly be determined from the extracted nonlinear features located close to the actuator. Therefore, we use $L_1$ regularization in this final layer with a strength of $10^{-6}$, which was found to give good results after tuning by hand. The architecture has a total of 3,512,308 trainable parameters.

The CNN needs to be calibrated with data that is representative of the distortions that will be observed on sky. One limitation is that the CNN can only be trained on wavefront profiles that are spanned by the DM modes. This means that higher spatial frequencies are not present in the training data, while they are observed on sky. Modelling the expected closed loop spatial power spectral density (PSD) and generating DM profiles approximating that distribution is an approach one could take. However, this would introduce a bias in the CNN. If the statistics of the residuals would differ from the data used to train the CNN this may degrade the reconstruction. Furthermore, the closed-loop PSD in itself depends on the reconstructor \cite{Jolissaint2006}. Due to the nonlinearity of the CNN this relation is not trivial. To introduce the least amount of bias in the reconstructor we have opted to train the CNN with independent normally distributed modal coefficients with a given standard deviation. To achieve good performance over a large range of wavefront errors the standard deviation of the amplitude distribution is sampled from a log-uniform distribution between 0.003 and 5 radians. In total, we simulate 80,000 input-output pairs. From this dataset, 80\% will be used for the training and 20\% for the validation and model selection. The loss function from equation \ref{eq:loss} is used with $\epsilon=0.01$ and we use the Adam algorithm with default initial hyperparameters as provided in \cite{Kingma2014} and a batch size of 64 to optimize the weights. To improve the accuracy of the model, we decay the learning rate by a factor two after every 20 epochs. Furthermore, the loss on the validation set is calculated after every epoch and the model is only saved when the validation loss decreases. Total training time of the model is $\sim$2 hours on a 12GB Tesla K80 GPU.

The interaction matrix is calibrated by applying DM modes with amplitudes of 0.3 rad, to ensure we are within the linearity range of the WFS. The same dataset as in the calibration for the CNN is used to find the optimal regularization strength for the MVM method, except only wavefronts with RMS below 0.3 rad are included. We iteratively test 20 different values of the regularization strength, logarithmically distributed between $10^{-5}$ and $10^5$ and the model with the lowest mean squared error will be used. For the $\beta=0$ profile we have to include larger amplitudes in the optimization of the regularization strength as 0.3 rad is too close to the sensitivity limit of the WFS for an input beam with $10^6$ photons. Since the wavefront sensor response is still linear over the considered range we include all data in the optimization.

\subsection{Simulation results: reconstruction of DM modes}
\begin{figure}[htbp]
    \centering
    \includegraphics[width=\linewidth]{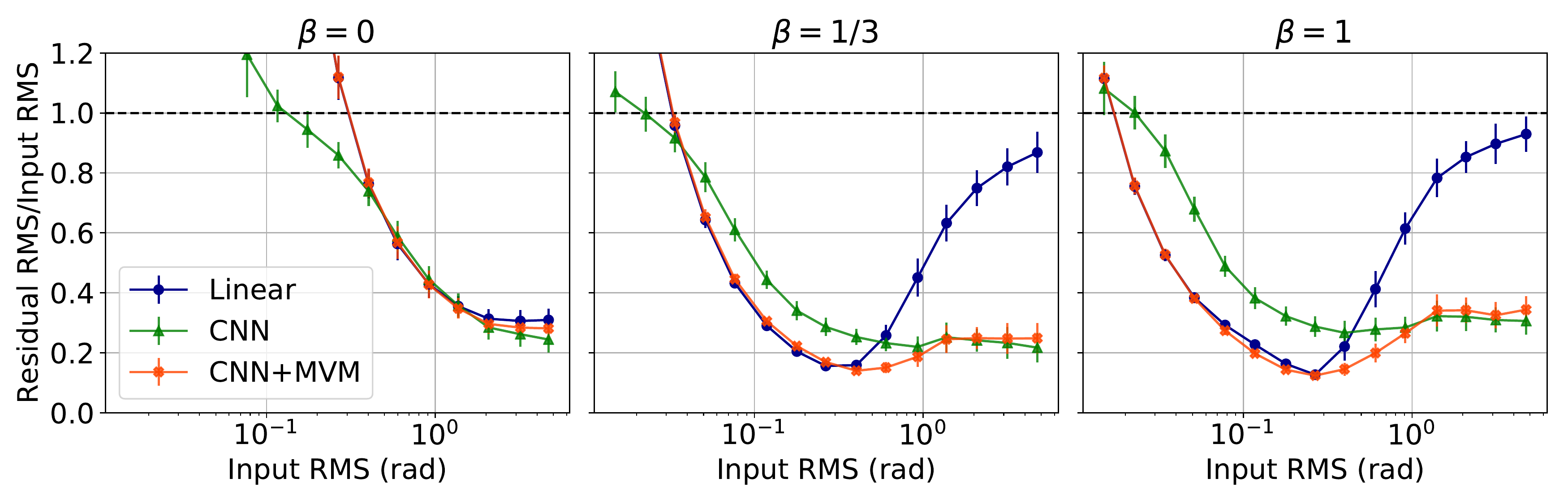}
    \caption{Comparison between the reconstruction methods for the simulated g-ODWFS. For each input RMS, 50 wavefronts are tested with normally distributed modal coefficients. All simulations use $10^6$ photons in the incoming beam.}
    \label{fig:simulations_reconstruction}
\end{figure}
We compare the reconstruction of the DM modes for the different reconstruction methods. Normally distributed modal coefficients with chosen standard deviation were applied on the DM and the calibrated models were used to reconstruct these coefficients. Figure \ref{fig:simulations_reconstruction} shows the relative residual root mean square phase variance (RMS) after the reconstruction for different step sizes in the rotation profile. For the profile without a step, we do not observe an increase in performance by using a CNN. This is as expected, since this profile has close to a linear response. The results for the $\beta=1/3$ and $\beta=1$ profiles show a significant improvement in the estimation of large phase aberrations by using a CNN. While the CNN-only reconstructor slightly degrades the estimation of small aberrations, the CNN+MVM approach does not have this problem while still providing similar improvement for large aberrations. Comparing the different step sizes shows the high sensitivity of the $\beta=1$ profile. However, this filter profile leads to more nonlinearities and has a smaller change in response in the nonlinear regime. This is shown by the increased residual RMS for both the linear and nonlinear reconstruction for large aberrations as compared to the filter with a step of 1/3. 

\subsection{Simulation results: closed loop performance}\label{section:closed_loop}
\begin{figure}[phtb]
    \centering
    \includegraphics[width=0.77\linewidth]{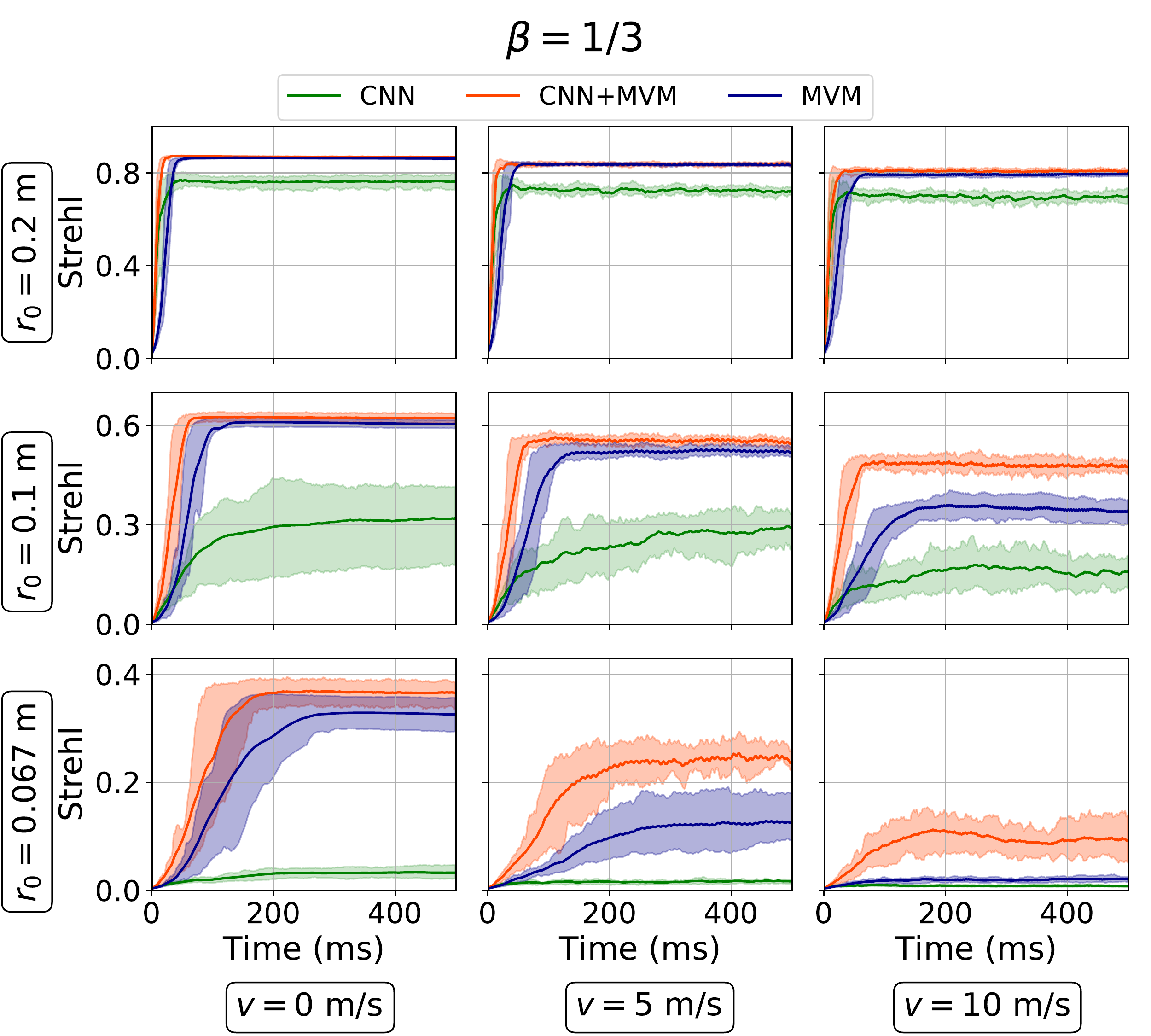}
    \includegraphics[width=0.77\linewidth]{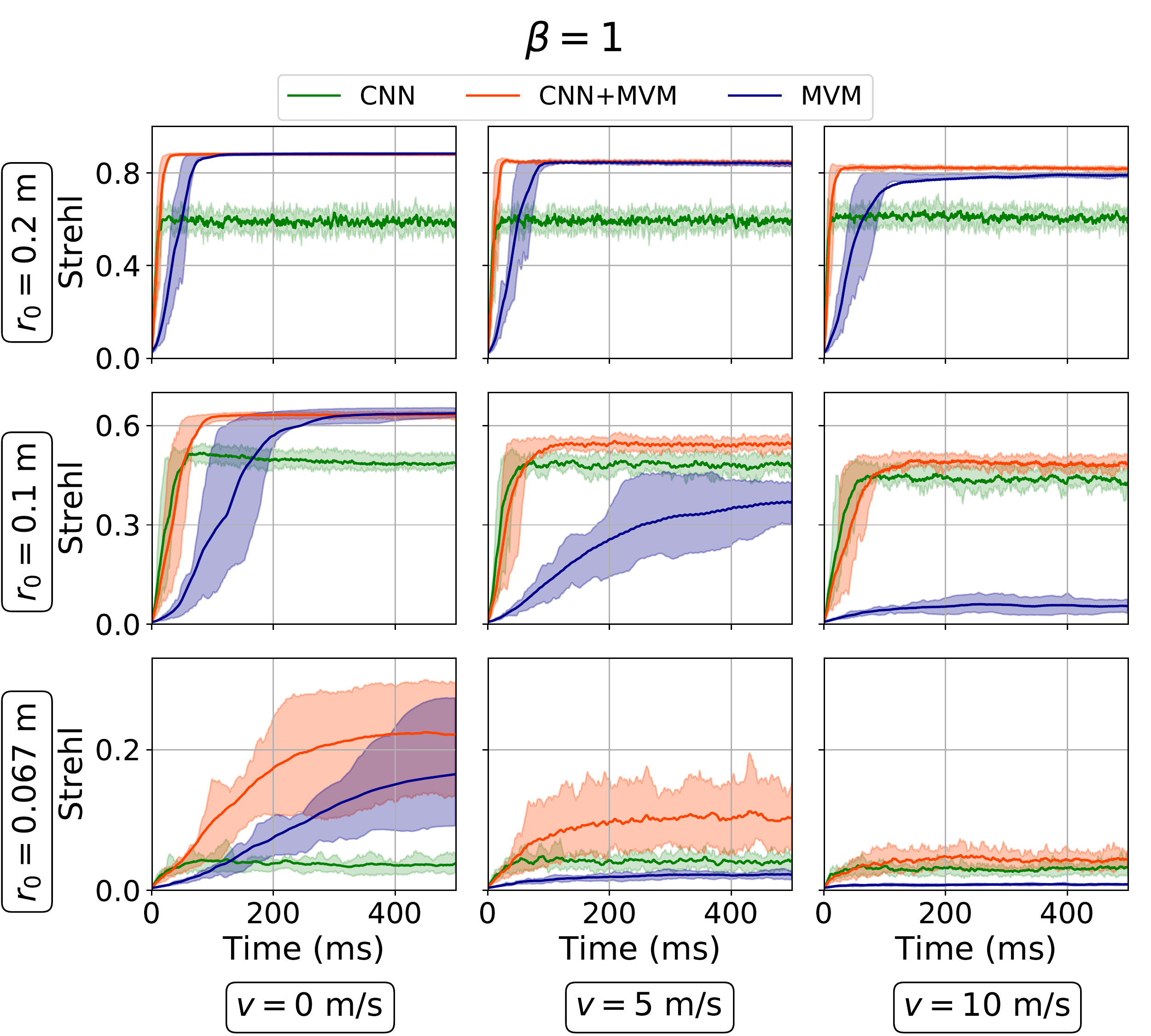}
    \caption{Results of the closed-loop simulations for different atmospheric conditions. Each plot shows the evolution of the Strehl ratio as a function of time for the given rotation profile, $r_0$ at 550 nm and wind velocity for the each of the considered reconstruction methods. The line shows the average Strehl ratio of 20 simulations while the coloured region shows the range between the 16th and 84th percentile. All simulations use photon noise levels for $10^6$ photons per WFS measurement.}
    \label{fig:closed_loop}
\end{figure}

The closed loop performance of the reconstruction methods is evaluated on a simulated single-layer atmospheric phase screen as implemented in hcipy with given Fried parameter $r_0$ and wind velocity $v$. The simulated AO system corrects at a frequency of 1 kHz and all simulations use a leaky integrator with a gain of 0.3 and leakage of 0.003. Figure \ref{fig:closed_loop} shows the results for the simulations under a variety of different atmospheric conditions. 

The simulations show that the CNN-only reconstructor does not perform well in closed loop situations. One explanation for this is that this is the result of higher order modes negatively influencing the reconstruction, as these are not present in the training data. Another explanation may be that the CNN gets stuck on a combination of modes that it cannot reconstruct. Due to the nonlinearity of the CNN there may be combinations of WFS measurements that exactly cancel out the output coefficients, which it may converge to in closed-loop. However, using the CNN as an additional nonlinear term on the MVM does not have this issue. We see that this approach leads to an increased Strehl ratio in the simulations with relatively bad conditions. Under these conditions the wavefront sensor operates in its nonlinear regime. In this case, the linear reconstructor underestimates the wavefront aberrations, while the CNN correction is able to improve the estimation. With the CNN the AO system is able to lock on to the highly variable turbulence, while the linear reconstruction can not do this. For larger filter step sizes $\beta$ the nonlinearity error increases. The extreme case of $\beta=1$ is used to demonstrate this. The $\beta=1$ filter is very similar to the unmodulated pyramid wavefront sensor. This shows that while the CNN is trained for the g-ODWFS, a CNN could also be used to improve the reconstruction of the unmodulated PWFS measurements. The resulting Strehl ratios from these simulations are shown in Table \ref{tab:strehl_ratios}.

\begin{figure}[htbp]
    \centering
    \includegraphics[width=0.96\linewidth]{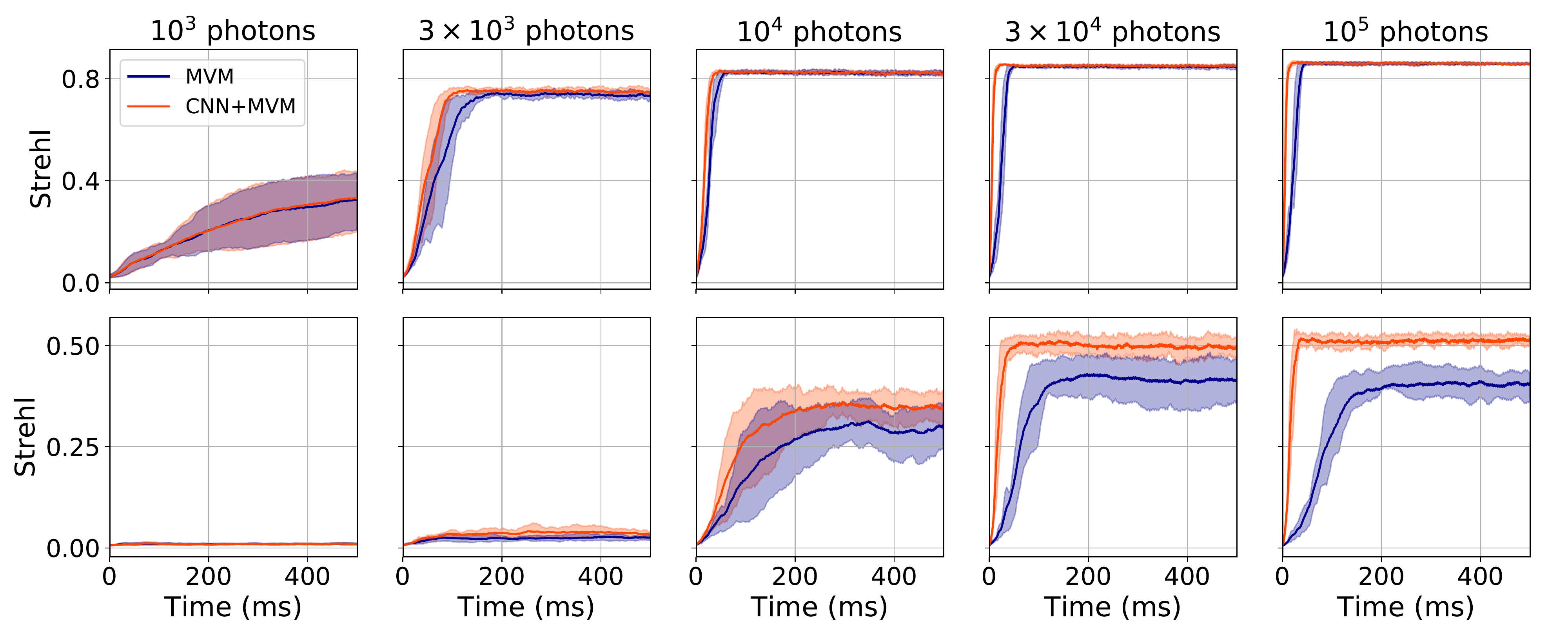}
    \caption{Results of closed loop simulations for different photon noise levels. The top row are simulations with conditions of $r_0=0.2$ m, $v=5$ m/s and the bottom row with $r_0=0.1$ m, $v=10$ m/s. All simulations use the filter profile with $\beta=1/3$. }
    \label{fig:closed_loop_photons}
\end{figure}

\begin{table}[htbp]
\caption{Strehl ratios (\%) obtained from the closed loop simulations. The values shown represent the mean Strehl from 20 simulations and errors are the standard deviation. Only the last 100 iterations are considered in the calculation.}
\label{tab:strehl_ratios}
\resizebox{\textwidth}{!}{%
\begin{tabular}{cc||ccc||ccc||}
\multicolumn{1}{l}{}                      & \multicolumn{1}{l||}{}        & \multicolumn{3}{c||}{$\beta = 1/3$}                                                  & \multicolumn{3}{c||}{$\beta = 1$}                                                   \\ \hline\hline
\multicolumn{1}{c|}{$r_0$ (cm)}              & \multicolumn{1}{c||}{v (m/s)} & \multicolumn{1}{c}{MVM} & \multicolumn{1}{c}{CNN} & \multicolumn{1}{c||}{CNN+MVM} & \multicolumn{1}{c}{MVM} & \multicolumn{1}{c}{CNN} & \multicolumn{1}{c||}{CNN+MVM} \\ \hline \hline

\multicolumn{1}{c|}{} &0& $86.2 \pm 0.5$ & $75.8 \pm 2.9$ & $86.6 \pm 0.4$ & $88.3 \pm 0.5$ & $58.8 \pm 1.6$ & $88.0 \pm 0.5$ \\ \hline
\multicolumn{1}{c|}{20} &5& $83.8 \pm 0.6$ & $72.2 \pm 1.1$ & $83.9 \pm 0.7$ & $83.9 \pm 0.9$ & $58.7 \pm 1.2$ & $84.3 \pm 1.1$ \\ \hline
\multicolumn{1}{c|}{} &10& $79.6 \pm 0.9$ & $68.9 \pm 2.8$ & $80.8 \pm 0.6$ & $79.1 \pm 1.2$ & $60.8 \pm 1.1$ & $81.9 \pm 0.8$ \\ \hline\hline
\multicolumn{1}{c|}{} &0& $60.4 \pm 1.3$ & $29.5 \pm 9.0$ & $61.9 \pm 1.4$ & $62.8 \pm 2.0$ & $49.6 \pm 2.0$ & $62.7 \pm 1.3$ \\ \hline
\multicolumn{1}{c|}{10} &5& $52.0 \pm 1.8$ & $27.4 \pm 5.7$ & $55.5 \pm 1.2$ & $33.9 \pm 8.4$ & $48.2 \pm 2.6$ & $53.9 \pm 2.1$ \\ \hline
\multicolumn{1}{c|}{} &10& $35.0 \pm 3.4$ & $15.8 \pm 4.6$ & $47.6 \pm 1.3$ & $5.3 \pm 2.6$ & $43.1 \pm 5.2$ & $48.2 \pm 3.2$ \\ \hline\hline
\multicolumn{1}{c|}{} &0& $33.2 \pm 2.7$ & $3.3 \pm 0.6$ & $37.3 \pm 1.9$ & $15.1 \pm 9.0$ & $3.8 \pm 1.5$ & $22.3 \pm 7.6$ \\ \hline
\multicolumn{1}{c|}{6.7} &5& $12.1 \pm 4.7$ & $1.4 \pm 0.3$ & $23.9 \pm 5.0$ & $2.2 \pm 0.6$ & $3.9 \pm 0.9$ & $11.7 \pm 5.1$ \\ \hline
\multicolumn{1}{c|}{} &10& $2.1 \pm 0.5$ & $0.7 \pm 0.1$ & $10.9 \pm 2.6$ & $0.8 \pm 0.1$ & $3.0 \pm 0.8$ & $4.5 \pm 1.2$ \\ \hline\hline

\end{tabular}}
\end{table}
\newpage
To test the effect photon noise has on the reconstruction, we have simulated the performance of the AO system for different photon fluxes. Figure \ref{fig:closed_loop_photons} shows the results of these simulations under conditions of $r_0=0.2$, $v=5$ m/s and $r_0=0.1$, $v= 10$ m/s respectively, all using the $\beta=1/3$ filter profile. The nonlinear correction does not provide any improved performance when there are not enough photon because the nonlinear structure is well within the noise. On the other hand, it also shows that the nonlinear correction does not lead to an amplification of the photon noise, as both methods have the same performance. When enough photons are available and the nonlinear residuals become visible, we again see the advantages of the increased effective dynamic range. This means that using the CNN as nonlinear correction only improves the reconstruction when enough photons are available for the nonlinear regime.

\section{Lab experiments}\label{section:Lab}
\subsection{Experimental setup}

The lab setup uses the AO system of the exoplanet path-finder instrument LEXI \cite{haffert2018sky}, which uses a g-ODWFS as its main wavefront sensor. To generate a diffraction-limited input beam a 633nm HeNe laser is injected into a single-mode fiber. A relay system with two achromatic doublets changes the focal ratio from F3 to F11. A 140 mm achromatic doublet collimates the beam. The system uses a Alpao 97-15 DM that is placed in the pupil plane of the collimated beam. A second achromatic lens with an identical focal length of 140 mm is used for 1 to 1 imaging. This image plane is the input of the g-ODWFS.

The g-ODWFS uses a 60 mm achromatic doublet to collimate the beam. A Wollaston prism is used to split the beam into two orthogonal polarizations. A 300 mm achromat is used to focus these two beams onto the focal plane mask. The focal plane mask is made of a single layer of patterned liquid-crystals\cite{miskiewicz2014direct,doelman2017patterned} and consists of two adjacent patterns with spatially varying fast-axes. The two patterns are used to measure the wavefront gradient in two directions. After the focal plane mask a second Wollaston prism, which is rotated by 45 degrees with respect to the first one, splits each beam into two again. This results in four output beams that are collimated by a third achromat with a 70 mm focal length onto an EM-CCD camera. The EM-CCD is conjungated to the DM surface. The DM surface is sampled with 45 pixels across the pupil.

The actuators of the DM are controlled in terms of their peak-to-valley amplitude. An amplitude of $\pm 1$ corresponds to the maximum actuator displacement of 30 $\mu$m in either direction. Data is taken for normally distributed actuator amplitudes with different variances. For each of the different amplitude variances used, 2000 random combinations of voltages are applied to the actuators and the images are saved. This gives a total of 16,000 measurements from which we use 60\% for training, 20\% for validation and 20\% for comparing the reconstruction.

The CNN architecture is slightly modified for the reconstruction of the lab data. The input array has a shape of 45x45x2 and since there are significantly less actuator modes we need to further reduce the spatial dimensions of the image. This is done by changing the stride of the first convolutional layer to 2. The fully connected layer now has 97 neurons, the number of actuators of the DM. The altered architecture has a total of 94,029 trainable parameters. The training procedure is the same as in the simulations but the loss function uses $\epsilon =10^{-5}$ due to the different units of the modal coefficients and the batch size is reduced to 8. The regularization strength for the MVM method is optimized using the data with mode RMS of up to 0.008. We have tested rotation profiles with steps of 0, 1/3 and 2/3. Since the simulations showed that the CNN-only approach is suboptimal, we will not consider it in this section.

\subsection{Lab results: reconstruction of DM modes}
\begin{figure}[htbp]
    \centering
    \includegraphics[width=\linewidth]{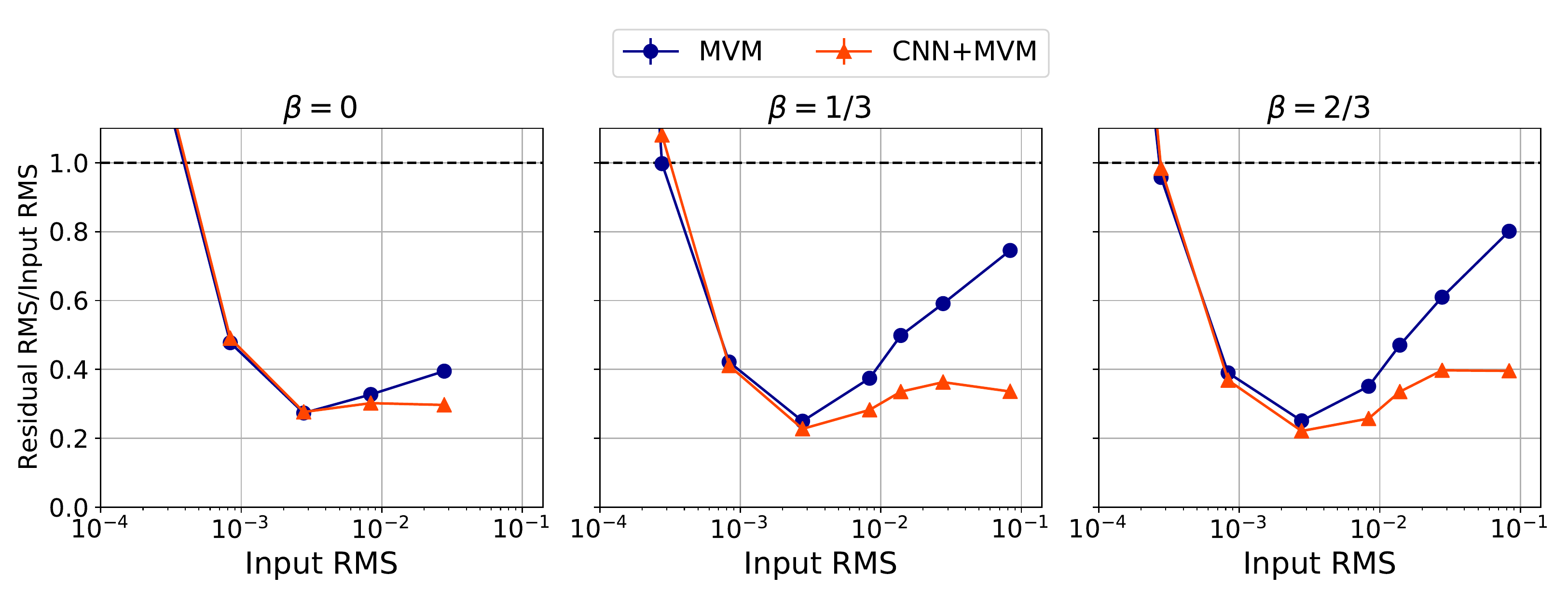}
    
    \caption{A comparison of the relative root mean square error of the residuals in DM units per actuator mode for the different reconstruction methods. Figures from left to right are for increasing step size $\beta$ in the rotation profile. }
    \label{fig:lab_results}
\end{figure}

\begin{figure}[htbp]
    \centering
    \includegraphics[width=0.75\linewidth]{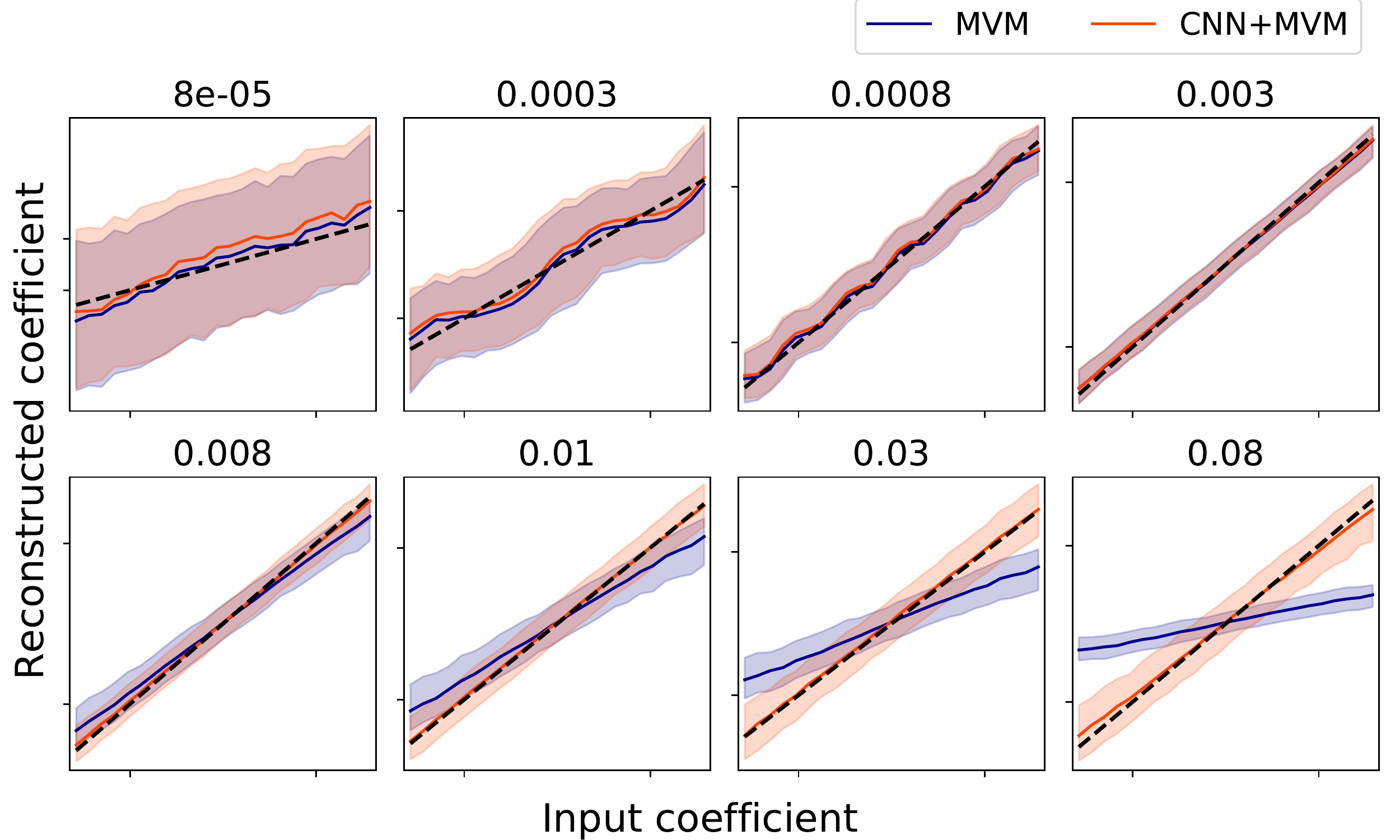}
    \caption{The plots show the median prediction of both methods for increasing mode RMS when the $\beta=1/3$ filter is used. The colored regions show the spread between the 16th and 84th percentile of the estimates. The dotted black line represents a perfect reconstruction. The title denotes the RMS in units of peak-to-valley amplitude.}
    \label{fig:lab_response}
\end{figure}

The performance of the reconstructors is evaluated on the test data and the results are shown in figure \ref{fig:lab_results}. This demonstrates that the CNN improves the reconstruction of larger modal coefficients. The reconstruction of these large aberrations is better for the $\beta=1/3$ rotation profile compared to $\beta=2/3$. The limit is not determined by the sensitivity of the WFS but due to other error sources. Figure \ref{fig:lab_response} shows the reconstructed coefficients as a function of the input coefficient for different input RMS with the $\beta=1/3$ rotation profile. This shows the underestimation of the coefficients by the linear model for larger aberrations, while the CNN is able to correct for the nonlinearities. These results confirm our findings from the simulations.

\section{Conclusions \& discussion}
In this paper, we have demonstrated the use of Convolutional Neural Networks for nonlinear wavefront reconstruction. Both simulations of the g-ODWFS and a lab implementation have shown an improvement in the estimation of large aberrations. Furthermore, we have compared the closed-loop performance of the reconstructors for simulated atmospheric turbulence. This showed that solely using a CNN results in suboptimal performance. On the other hand, using the CNN as a nonlinear correction term on the MVM results in a higher Strehl ratio under conditions where the WFS operates in its nonlinear regime as compared to the standard MVM approach. It was demonstrated that the nonlinear correction only provides improved effective dynamic range when the number of photons is sufficiently high such that nonlinear structures can be measured. The results show that the CNN does not amplify noise, which is a concern for nonlinear reconstruction methods.


The nonlinear correction does come at additional computational cost, often measured in the number of floating point operations (FLOPs). For the architecture used in the simulations propagating a single measurement through the CNN takes $\sim 73$ megaFLOPs compared to $\sim 7 $ megaFLOPs for the MVM. For the lab architecture this reduces to $\sim 9$ megaFLOPs and $\sim 0.4$ megaFLOPs respectively. The cost for the CNN+MVM approach is the sum of these two. Astronomical AO systems run at roughly 1 kHz, so to reach these speeds the algorithm requires a computer with the computational power of at least 100 GFLOPs per second, which is well within the current computing capabilities. There are a few possibilities for decreasing the computational demand in our approach. First of all, a lot of research has been done on creating efficient Convolutional Neural Network architectures for low latency predictions on mobile hardware \cite{Howard2017,Freeman2018}. We did not focus on efficiency and it should be relatively easy to decrease the computational load by optimizing the architecture. Secondly, the matrix multiplications in the prediction can be heavily parallelized on a GPU. Finally, one could consider reconstructing less nonlinear modes. However, the wavefront sensor measurements will still need to be propagated through the CNN, even if the output dimension is smaller. Applying software binning on the wavefront sensor measurements may relieve some of the computational demand but may also increase the effect of higher order modes on the reconstruction.

Our work shows that there is still much to be gained by improving the wavefront reconstruction algorithms, especially in the nonlinear regime. By applying nonlinear corrections we are able to reach equal or higher Strehl ratios under all atmospheric conditions. In some cases the gain in Strehl is more than 40 percentage points. Adding these nonlinear corrections could be important for the upcoming Extremely Large Telescopes that will need AO to operate their instruments. The precise impact is difficult to estimate because the nonlinearity error depends strongly on the amount of turbulence cells across the pupil. All ELTs have a significantly larger aperture than the one that was considered in this work and the instruments that will use a PWFS all sense the wavefront at different wavelengths. Both aspects influence the nonlinearity error, and make the precise impact instrument specific. It is therefore important both to test the proposed reconstruction method at the telescope, to show that it can operate under realistic turbulence, and to apply the proposed technique with accurate models for the ELT-sized AO instruments.

\section*{Acknowledgements}
The authors would like to thank Prof. C.U. Keller for the supervision of the research and Dr. M. Kenworthy for the feedback on this work. We also thank the anonymous reviewers for their comments to improve this work.

\section*{Disclosure}
The authors declare no conflicts of interest.

\bibliography{main}


\end{document}